
\nopagenumbers
\font\ninerm=cmr9
\font\fourteenrm=cmb10 scaled\magstep 2
\font\bbf=cmb10 scaled\magstep 1
\font\greekbold=cmmib10
\def\title#1{\bigskip\noindent{\bbf #1}\hfill\rm\medskip}
\vsize=19.3 cm
\hsize=12.2 cm\noindent
\baselineskip=12pt
\hoffset=2.4 cm
\voffset=1.8cm

\noindent\hfill
YCTP-P1-95
\bigskip
\noindent{\fourteenrm On Abelian Bosonization of Free Fermi
Fields\hfill

\noindent in Three Space Dimensions}
\vglue 1 cm
\noindent
Ramesh Abhiraman and Charles M.
Sommerfield\footnote*{\ninerm\noindent Based on a talk given by
C. Sommerfield at G\"ursey Memorial Conference I, Bo\v gazi\c ci
University, Istanbul, Turkey, June 1994.}
\bigskip
\noindent {\ninerm Center for  Theoretical Physics, Yale
University, New Haven, Connecticut, USA}
\bigskip
\bigskip
\vglue 1.1cm
\noindent{\bf Abstract.} One of the methods used to
extend two-dimensional bosonization to four
space-time dimensions involves a
transformation to new spatial variables so
that only one of them appears
kinematically.  The problem is then reduced
to an Abelian version of two-dimensional
bosonization with extra ``internal''
coordinates.  On a formal level, putting
these internal coordinates on a finite
lattice seems to provide a well-defined
prescription for calculating correlation
functions.  However, in the
infinite-lattice or continuum limits,
certain difficulties appear that require
very delicate specification of all of the
many limiting procedures involved in the
construction.
\title{Introduction}
\noindent After the initial success of
bosonization in 1+1 dimensions [1,2] Alan Luther presented a
heuristic formula [3] for bosonization of free, massless
relativistic fermions in 3+1 dimensions.   A  configuration-space
transform [4] was later used  by H. Aratyn [5] to put this into a
prettier but no less heuristic form.
For other approaches to this problem the reader is referred to
References [6] and [7].
\smallskip
The basic idea of the method described
here is to view field theories in 3+1 dimensions as 1+1
dimensional theories with an internal, non-kinetic degree of
freedom.  The main subject of this paper is to examine carefully
the limiting procedures involved in bosonizing a 1+1 dimensional
free fermion with an internal degree of freedom that takes
values that are in a continuum.
\title{Tomographic transform}
\noindent We begin by reviewing the transformation to
a 1+1 dimensional theory both for bosons
and for fermions.  We then present the standard Abelian
technique  for bosonizing the 1+1 dimensional fermion when the
internal degrees of freedom lie on a finite lattice.  There
follows  a careful  analysis of the continuum limit of this
lattice in the context of a simplified model.
\smallskip
The  tomographic transform of a
free scalar field  $\phi({\bf x},t)$ with mass $m$ in 3+1
dimensions is defined as
$$
\tilde\phi(y,{\bf n},t)={1\over2\pi}\int
d^3{\bf x}\; \partial_y\delta(y-{\bf n}\cdot{\bf x})
\phi({\bf x},t).
$$
Here ${\bf n}$ is a unit
vector and
$d^2{\bf n}$ an element of solid angle in the direction
of ${\bf n}$.
\smallskip
The field equation satisfied by $\tilde\phi(y,{\bf n},t)$ and its
equal time commutation relations are essentially those of a 1+1
dimensional scalar field, also free and with mass $m$.
\smallskip
The corresponding transform for a fermion field is
$$\tilde\psi^a(y,{\bf
n},t)={1\over2\pi}
\int d^3{\bf x}\;\partial_y\delta(y-{\bf n\cdot x})
 \sum_\alpha u^{\dagger a}_\alpha({\bf
n})\psi_\alpha({\bf x},t).
$$
It satisfies the 1+1 dimensional Dirac equation for a
right-moving  fermion field, (that is, depending only on $y-t$) and
has the expected equal-time anticommutation relations. Here
$u^a({\bf n})$, $a=1,2$, are orthonormal four-component spinors
that satisfy
$(\hbox{\greekbold\char'013}\cdot{\bf n})u^a({\bf
n})=u^a({\bf n})$ where \hbox{\greekbold\char'013} are the Dirac
alpha matrices.
\smallskip
The strategy is to use standard Abelian 1+1 dimensional
bosonization to construct the right-moving  $\tilde\psi^a(y,{\bf
n},t)$, in terms of the
right-moving part of the transformed boson
$\tilde\phi(y,{\bf n},t)$, assuming, to begin with, that
${\bf n}$ is restricted to a lattice.
This is given formally, with the  $t$ dependence suppressed, as
$$\tilde\phi_r^a(y,{\bf n})=
\hbox{$1\over2$}[\tilde\phi^a(y,{\bf n})-
\hbox{$1\over2$}\int_{-\infty}^\infty dy'\;
\epsilon(y-y')\,\partial_t\tilde\phi^a(y',{\bf n})],
$$
where $\epsilon(y-y')=(y-y')/|y-y'|$. It
follows that
$$
\langle0|\tilde\phi_r^a(y,{\bf n})
\tilde\phi_r^b(y',{\bf
n'})|0\rangle=-{1\over4\pi}\delta^{ab}\delta({\bf n},{\bf
n}')
\ln\left(\mu[\alpha-i(y-y')]\right)
$$
where $\mu$  and $\alpha$ are infrared and ultraviolet
cutoffs, respectively.
\smallskip
We first take ${\bf n}$ to be discretely
distributed on some lattice of directions
and reinterpret the Dirac delta function  as a Kronecker delta
function.
Replacing the pair $a$ and
${\bf n}$ by a single index $A$, we find the formal
bosonization expression
$$\tilde\psi^A(y)=
{1\over\sqrt{2\pi\alpha}}e^{i\tilde\phi_r^A(y)} K^A$$ where
$$K^A=\exp{i\over2}\sqrt{\pi}\sum\nolimits_C
\epsilon^{AC}[\phi^C_r(\infty)-\phi_r^C(-\infty)]
$$
and where $\epsilon^{AB}=-\epsilon^{BA}$, with
$|\epsilon^{AB}|^2=1$. The Klein factors $K^A$ are needed to
make sure that fermion fields with different indices anticommute
rather than commute.  In what follows we will ignore these
factors.  Their only effect is to correct an occasional sign to
its proper value.
\smallskip
The correlation functions for the fermion field are
determined from the matrix elements of their bosonic
representation
in the bosonic vacuum. The
two-point function is computed as
$$
\langle0|\tilde\psi^A(y)\tilde\psi^{\dagger
B}(y')|0\rangle
={1\over2\pi}{\mu\over\{\mu[\alpha+i(y-y')]\}
^{\delta^{AB}}}.
$$
If $A=B$ we get the known fermion function.
If $A\ne B$ then this vanishes in the limit
$\mu\rightarrow0$, so that the result is indeed
proportional to $\delta^{AB}$.
When reinterpreted in terms of a continuous ${\bf n}$ we do get
the correct two-point function for the transformed fermion
field:
$$\langle0|\tilde\psi^a(y,{\bf n})
\tilde\psi^{\dagger b}(y',{\bf n}')|0\rangle
={1\over2\pi}\;{\delta^{ab}\delta({\bf
n},{\bf n}')\over\alpha-i(y-y')}.
$$
 All of the other fermion
correlation functions come out properly as well.
\smallskip
To test the consistency of the bosonization
formula we must consider the important fermion
bilinear operators in the 3+1 dimensional
theory such as the Poincar\'e group
generators and the chiral charge operators.
If $\Omega$ is one of these, does
 the sequence
$
\Omega[\psi_{3+1}]\leftrightarrow
\Omega[\tilde\psi_{1+1}]\leftrightarrow
\Omega[\tilde\phi_{1+1}]\leftrightarrow
\Omega[\phi_{3+1}]
$
make sense?
Trouble arises when we consider the rotation and boost
generators. The proper treatment of the continuum
limit of the lattice plays a fundamental role here.
\title{Simplified model and smoothing functions}
\noindent In order to get at the heart of the problem of the
continuum limit we will treat a simple case in which the the
internal variable is one-dimensional.  We thus  consider a single
component 1+1 dimensional massless chiral fermion field
$\psi(x,u)$ that depends on a continuous internal variable
$u$ whose domain is the real line. The bosonization of the
fermion field in the Dirac equation
$(\partial_t+\partial_x)\psi(x,u)=0$ will be in terms of a
right-moving chiral massless boson field $\phi_r(x,u)$.  Assuming
for the moment that $u$ is a discrete variable,  we write, up to
Klein factors,
$\psi(x,u)=(2\pi\alpha)^{-1/2}
\exp{[i\sqrt{4\pi}\phi_r(x,u)]}$
and obtain for the fermion two-point function
$$
\langle0|\psi(x,u)\psi^{\dagger}(x',u')|0\rangle={1\over
2\pi}{\delta_{u,u'}\over \alpha-i (x-x')}$$
\smallskip
For the bosonization procedure to work, it is clear that the
exponent that appears in the evaluation of the fermion two-point
function must be associated with a Kronecker delta function,
rather than a Dirac delta function so that the logarithmic
singularity in the bosonic two-point function exponentiates to a
simple pole or a simple zero.  We need a more analytic method of
converting one type delta function to the other.  Returning to a
continuous
$u$, we introduce a new boson field $\Phi(x,u)
=\int du' f(u-u')\phi_r(x,u').
$
We take $f(u)$ to be a real, even function.  The Dirac delta
function is replaced in the commutation relations among the
$\Phi$ operators, as well as in the bosonic two-point function, by
the convolution $g(u-u')=\int du''\;f(u-u'')f(u''-u')$. It is
thereby softened.  We will choose
$f(u)$ so that $g(u)\ge 0$, $g(0)=1$, and $\epsilon(u)g'(u)<0$
if
$g(u)\ne0$.  To make sure that $g(u)$ goes to 0 very
quickly as $|u|$ increases we take it to be a
function of $u/\lambda$ and eventually take the scale parameter
$\lambda$ to zero. This simulates the properties of a Kronecker
delta function.
\smallskip
 We define  the field $\hat\psi(x,u)=
C(2\pi\alpha)^{-1/2}\exp{[i\sqrt{4\pi}\Phi(x,u)]}$ which will be the candidate
for the canonical fermion field $\psi$ in the appropriate limit.
Here $C$ is some constant, to be specified later, that
depends on $\mu$ and on the choice of $g(u)$.  We then find for the
fermionic two-point function
$$\langle0|\hat\psi(x,u)
\hat\psi^\dagger(x',u')|0\rangle
=|C|^2\mu^{[1-g(u,u')]}{1\over
2\pi}{1\over[\alpha-i(x-x')]^{g(u-u')}}.
$$
As the infra-red cutoff $\mu$ goes to zero, this vanishes
unless $u=u'$.  Choosing $C$ to depend on $\mu$ so that $C$
diverges suitably in this limit, we obtain the desired Dirac delta
function and the same answer as in the discrete case, with the
Kronecker delta replaced by the Dirac delta.
\smallskip
There is a surprise when we consider the four-point function.
One must analyze the behavior of the quantity
$|C|^4\mu^{(2+g_{12}+g_{34}-g_{13}-g_{23}-g_{14}-g_{24})}$ as
$\mu\rightarrow0$.  Here $g_{ij}=g_(u_i-u_j)$ with $u_1$ and
$u_2$ assigned to the fermion fields and $u_3$ and $u_4$ to the
adjoint fields.  In order to obtain the correct behavior when
all four internal indices are close we find that we cannot
choose $f(u)$ such that $g'(0)=0$.  In fact $g(u)$ must have
a cusp at $u=0$.  Thus Gaussian functions for $f(u)$ and
$g(u)$ are ruled out.  A triangular pulse is acceptable for
$g(u)$ and this corresponds to a rectangular pulse for
$f(u)$.
\title{Fermion bilinears}
\noindent We go on to discuss the fermion bilinears. This is where
we had trouble in the discrete case.  A generic form of such a
bilinear operator is
$[\psi^\dagger(x_{2},u_2),\psi(x_{1},u_1)]$ where the arguments
are allowed to come together, possibly after various
derivatives have been taken.  This is the
nature of the spatially point-split structure of the charge
operator, and of the generators of translations in $x$, $t$ and
$u$.  We proceed to evaluate it using operator-product expansion
methods (which are exact in this model).  We have
$$
[\hat\psi_2^\dagger,\hat\psi_1]
=-{|C|^2\over2\pi}
\mu^{(1-g_{12})}(R_{12}{}^{-g_{12}}-
R_{21}{}^{-g_{21}})
\hbox{\bf:}e^{i\sqrt{4\pi}(\Phi_1-\Phi_2)}\hbox{\bf:}.
$$
Here $R_{12}=\alpha-i(x_1-x_2)$.  The normal ordering is with
respect to the bosonic creation and annihilation operators.  We
introduce
$u={1\over2}(u_1+u_2)$,
$v=u_1-u_2$,
$x={1\over2}(x_1+x_2)$ and
$\xi=x_1-x_2$, and obtain
$$
[\hat\psi_2^\dagger,\hat\psi_1]
=-{|C|^2\over2\pi}
\mu^{[1-g_{12}]}
[R(\xi)^{-g(v)}-R(-\xi)^{-g(v)}]\Omega(\xi,v)
$$
where
$\Omega(\xi,v)=\;\hbox{\bf:}\exp i\sqrt{4\pi}[
\Phi(x+{1\over2}\xi,u+{1\over2}v)
- \Phi(x-{1\over2}\xi,u-{1\over2}v)]\hbox{\bf:}$.  Taking $v$
and $\xi$ small, (while keeping  $\alpha\ll\xi$),  and sending
$\mu$ to 0 we discover that
$[\hat\psi_2^\dagger,\hat\psi_1]\rightarrow-i\delta(v)
\Omega(\xi,0)/(\pi\xi)$.
\smallskip
The chiral charge operator $Q$ is
given in terms of the canonical fermions by
$$
\hat Q =\hbox{$1\over2$}\int dx\;du_1\;du_2\; \delta(u_1-u_2)
[\psi^\dagger(x,u_2),\psi(x,u_1)].$$
We choose to smooth out the $u$ delta function and replace it by
$N_\lambda g(u_1-u_2)$ where $N_\lambda$ is chosen so that
$N_\lambda\int du_1\;g(u_1-u_2)=1$.   Then $N_\lambda g(u_1-u_2)$
becomes a Dirac delta function as
$\lambda\rightarrow0$.
We construct the candidate chiral charge $\hat Q$
in terms of $\hat\psi$ as
$\hat Q=\lim_{\xi\rightarrow0}N_\lambda\int
dx\;du\; dv
\;g(v){1\over2}[\hat\psi^\dagger_2,\hat\psi_1]$ so that
$$Q=\lim\limits_{\xi\rightarrow0}{-iN_\lambda g(0)\over4\pi
\xi} \int dx\;
du\;[\Omega(\xi,0)-\Omega(-\xi,0)]={N_\lambda\over
\sqrt{\pi}}\int dx\; du\;{\partial\over\partial
x}\Phi(x,u).
$$
The charge density in $u$ space is thus $\hat Q(u)=N_\lambda
\pi^{-1/2}[\Phi(\infty,0)-\Phi(-\infty,0)]$.
We then have $[\hat \psi(x,u),\hat Q(u')]=N_\lambda
g(u-u')\hat\psi(x,u)$ which becomes the correct canonical
relation as
$\lambda\rightarrow0$.
\smallskip
The canonical operator
$J=-i{1\over4}\int dx\; du\;[\psi^\dagger(x,u),
{\partial\over\partial u}\psi(x,u)]+(\hbox{h. c.})$
 generates translations in $u$.  It is
the analogue of the  rotation and boost operators in the
tomographic representation of the 3+1 dimensional case.
If we were to follow the procedure used for the chiral
charge we would find that when expressed in terms of
bosonic fields, the candidate generator would vanish.
The remedy for this is to back up a step and delay taking
$\mu$ to zero.  Then the spreading function used
to define the bilinear, (which need not be the same as what
 we used for the chiral charge, namely $N_\lambda g(u)$), can be
taken to depend on $\mu$ in just such a way as to yield
the correct answer in terms of canonical boson fields
in the limit.
\bigskip
\noindent {\bf Acknowledgment.}  This  work was partially supported
by the United States Department of Energy under grant
DE-FG02-92ER 40704.
\bigskip
\noindent{\bbf References}\hfill\ninerm\medskip\noindent
\noindent\item{[1]} S.~Coleman, {\sl Phys.~Rev.~D\/} {\bf 11}
(1975) 2088.
\smallskip
\noindent\item{[2]} S.~Mandelstam, {\sl Phys.~Rev.~D\/} {\bf 11}
(1975) 3026.
\smallskip
\item{[3]} A.~Luther, {\sl Phys.~Rev.~B\/}
 {\bf 19} (1979) 320.
\smallskip
\item{[4]} C.~M.~Sommerfield,  in ``Symmetries in
Particle Physics,'' A.~Chodos, I.~Bars and C.-H.~Tze, eds.,
Plenum Press (1984) pp.~127-140.
\smallskip
\item{[5]} H.~Aratyn, {\sl Nuc.~Phys.~B\/}
 {\bf 227} (1983) 172.
\smallskip
\item{[6]} E.~C.~Marino, {\sl Phys.~Lett.\/} {\bf B263}
(1991) 63.
\smallskip
\item{[7]} A.~Kovner and P.~S.~Kurzepa, {\sl
Phys.~Lett.\/} {\bf B328} (1994) 506.
\end